\documentclass{emulateapj}

\slugcomment{Submitted to the Astrophysical Journal}
\shorttitle{The Star Formation-Density Relation}
\shortauthors{Quadri et al.}

\begin{document}

\title{Tracing the Star Formation-Density Relation to $z \sim 2$}

\author{
Ryan F.~Quadri\altaffilmark{1,2,3},
Rik J.~Williams\altaffilmark{1},
Marijn Franx\altaffilmark{3},
Hendrik Hildebrandt\altaffilmark{4}}

\altaffiltext{1}{Carnegie Observatories, Pasadena, CA 91101}
\altaffiltext{2}{Hubble Fellow}
\altaffiltext{3}{Leiden Observatory, Leiden University, NL-2300 RA,
  Leiden, Netherlands}
\altaffiltext{4}{University of British Columbia, Vancouver, B.C. V6T 2C2,
Canada}
\email{quadri@obs.carnegiescience.edu}

\begin{abstract}
Recent work has shown that the star formation-density relation --- in
which galaxies with low star formation rates are preferentially found
in dense environments --- is still in place at $z \sim 1$, but the
situation becomes less clear at higher redshifts.  We use
mass-selected samples drawn from the UKIDSS Ultra-Deep Survey to show
that galaxies with quenched star formation tend to reside in dense
environments out to at least $z \sim 1.8$.  Over most of this redshift
range we are able to demonstrate that this star formation-density
relation holds even at fixed stellar mass.  The environmental
quenching of star formation appears to operate with similar efficiency
on all galaxies regardless of stellar mass.  Nevertheless, the
environment plays a greater role in the build-up of the red sequence
at lower masses, whereas other quenching processes dominate at higher
masses.  In addition to a statistical analysis of environmental
densities, we investigate a cluster at $z = 1.6$, and show that the
central region has an elevated fraction of quiescent objects relative
to the field.  Although the uncertainties are large, the environmental
quenching efficiency in this cluster is consistent with that of galaxy
groups and clusters at $z \sim 0$.  In this work we rely on
photometric redshifts, and describe some of the pitfalls that large
redshift errors can present.
\end{abstract}
\keywords{galaxies: evolution -- galaxies: high-redshift -- galaxies:
  clusters: general -- cosmology: large-scale structure of the
  universe}

\section{Introduction}
\label{sec:introduction}

One of the most striking features of the galaxy population at low
redshift is the correlation between galaxy properties --- notably,
mass, morphology, and star formation rate --- and the local
environmental density
\citep[e.g.][]{dressler80,balogh98,gomez03,kauffmann04,weinmann06}.  Although
significant work has gone into understanding these relationships from
both an observational and a theoretical point of view, we still have a
limited understanding of the role of environment in shaping galaxy
properties and of the specific physical processes through which the
environment operates.  In this context, it may be expected that
extending the obervations to higher redshift will provide new insight.

In this paper we investigate the evolution of the star formation-
density relation, in which galaxies in dense regions tend to have
lower star formation (SF) rates than galaxies in the field.  A number
of studies in recent years have used statistical measures of
environmental densities to investigate this very issue, and have
mostly concluded that the SF-density relation disappears, or even
``reverses,'' at $z \sim 1$ \citep[e.g.][]{cucciati06, cooper07,
  elbaz07, ideue09, salimbeni09, scodeggio09, tran10, grutzbauch11a}.
But several of these studies do find robust environmental effects out
to $z \sim $0.8--1 \citep[see
  also][]{scoville07,patel09b,patel11,cooper10}, which implies a
fairly sharp transition at $z \sim 1$.

However there are several reasons to doubt the existence of such a
strong transition.  Given the apparently smooth growth of the red
sequence over cosmic time \citep{brammer09, williams09, ilbert10,
  kajisawa11}, it would be odd if this growth occurred preferentially
in overdense regions at $z < 1$ but avoided them at $z > 1$.  Instead,
there are several examples of clusters at $z \sim 1.5$ that already
have prominent populations of passive galaxies in place
\citep[e.g.][]{mccarthy07,kurk09,wilson09,strazzullo10}.  Studies of
galaxy clustering have also found that red galaxies tend to be more
clustered than blue galaxies at $z \gtrsim 1.5$
\citep[e.g.][]{grazian06,quadri07,quadri08,hartley10}, which also
suggests that they reside in denser environments.  Moreover, if
nothing else, it is natural to expect that more massive galaxies at $z
> 1$ should tend to lie in denser environments, and given that more
massive galaxies are also more likely to have their star formation
quenched (``downsizing''), this would suggest that the SF-density
relation should extend to $z > 1$.

Several of the studies that have performed direct estimates of
environmental densities have used large samples of galaxies with
spectroscopic redshifts, however obtaining spectroscopic redshifts for
large and unbiased samples at $z > 1$ is very difficult.  Quiescent
galaxies, which are very faint in the observer's optical, are
especially hard to observe, and it is these galaxies that are of
particular interest when studying the SF-density relation.
A second, although perhaps less significant, difficulty for spectroscopic
studies is obtaining sufficiently dense spectroscopic sampling of
galaxies in overdense regions.

Other studies have estimated environmental densities using photometric
redshifts.  This has the obvious advantage of providing less biased
samples, and is also not subject to the geometrical constraints
imposed by multi-object spectrographs.  However these studies have the
tremendous disadvantage of large redshift errors, making it virtually
impossible to confirm the membership of any one individual galaxy in
an observed overdensity, or to know whether an apparent overdensity is
a single structure or simply arises in projection.  But the hope is
that by using a large enough sample it will be possible to tell
whether galaxies of one type tend to have, on average, a greater or
fewer number of near neighbors with similar photometric redshifts than
galaxies of another type.

In this paper we wish to study the evolution of the SF-density
relation out to $z \sim 2$ using mass-selected samples, as this is
expected to give a more representative view of environmental effects
than flux- or luminosity-selected samples \citep[e.g.][]{patel09b}.
Because many of the most massive galaxies, and in particular those
with quenched star formation, are very faint in the observer's
optical, it is not feasible with current telescopes to obtain large
spectroscopic samples.  This forces us to rely on photometric
redshifts.

We separate the quiescent from the star forming galaxies using the
observed bimodality in a color-color diagram, which is
model-independent and has a significant advantage over the traditional
color-magnitude diagram in that it removes contamination of the ``red
sequence'' by dusty star-forming galaxies.  The SF-density relation is
then quantified using the fraction of quiescent galaxies in different
environments.  The dataset is described in \S \ref{sec:data}, and in
\S \ref{sec:density_estimators} we discuss how we estimate the
environmental densities, taking care to note the ways in which the use
of photometric redshifts can actually introduce artificial
environmental trends.  The results are presented in \S
\ref{sec:rfrac}, and in \S \ref{sec:cluster} we consider the
particular case of a cluster at z=1.62.  Conclusions are given in \S
\ref{sec:summary}.  In the Appendix we discuss some additional details
of the relationships between stellar mass, star formation, and
environmental density, and we demonstrate the relative importance of
environmental processes in the build-up of the red sequence.

\section{Data and Object Selection}
\label{sec:data}

\subsection{Imaging data}

Here we use public data in the field covered by the UKIDSS Ultra-Deep
Survey (UDS; O.~Almaini 2011, in
preparation).\footnote{www.nottingham.ac.uk/astronomy/UDS/} This field
has extensive imaging in the optical, near-infrared (NIR), and
infrared (IR).  The NIR data comes from the UKIDSS survey, which is
described by \citet{lawrence07}.  The photometric system and
calibration are described by \citet{hewett06} and \citet{hodgkin09},
respectively.  Optical $BVRi'z'$ imaging in this field comes from the
Subaru/\emph{XMM} Deep Survey \citep{furusawa08}, with additional
$u^*$-band taken with MEGACAM on the Canada-France Hawaii Telescope
(P.I.~O.~Almaini).  We use \emph{Spitzer}/IRAC and MIPS photometry
from the \emph{Spitzer}-UDS Survey (SpUDS; P.I.~J.~Dunlop).  The IR
fluxes were measured using the PSF-convolution procedure of
\citet{labbe06}.  The area of the field with full multiwavelength
imaging is $\sim$0.65$\rm{deg}^2$.

We update our previous photometric $K$-selected catalog of the UDS
\citep[described in][]{williams09} using Data Release 8 of the UKIDSS
NIR imaging, which reaches $5\sigma$ point-source depths of $J=24.9$,
$H=24.1$, and $K=24.5$ (AB magnitudes).  We also update the optical
photometry using SXDS Data Release 1, and use the SpUDS IR data rather
than the shallower data from the \emph{Spitzer} Wide-Area Infrared
Extragalactic Survey \citep[SWIRE;][]{lonsdale03}.  This updated
catalog will be described in detail by Williams et al.~(2011; in
preparation).

\subsection{Photometric redshifts}
\label{sec:zphots}

We calculate photometric redshifts with EAZY \citep{brammer08}.  We
make use of a slightly updated template set (G.~Brammer, private
communication), and use an iterative zeropoint-tuning procedure which
is effective at removing systematic errors in the photometric
redshifts.  A comparison to a sample of $\sim1500$ spectroscopic
redshifts drawn from a variety of sources \citep[Simpson et al. 2010,
  in preparation; Akiyama et al. 2011, in preparation;][]{smail08}
suggests a typical uncertainty in $\Delta z/(1+z)$ of 0.018 over
$0<z<1$, and of 0.022 over $1<z<1.5$.  An additional comparison to the
spectroscopic redshifts of objects in the z=1.6 cluster in the UDS
\citep{papovich10, tanaka10} gives an uncertainty of $\sim 0.03$.

While this analysis suggests that the photometric redshift quality is
very good, the spectroscopic redshifts that we have compared to are
not representative of the galaxies studied in this paper, so exact
values quoted above may not be particularly relevant.  And, as
discussed in \S \ref{sec:density_estimators}, overestimating the
quality of the photometric redshifts can introduce spurious signals in
environmental studies.  We therefore use the procedure of
\citet{quadri10} to estimate the true redshift errors: briefly, close
pairs of objects on the sky have a significant probability of lying at
the same redshift, so the photometric redshift differences in close
pairs can be used to estimate the distribution of photometric redshift
errors.  A statistical correction for close pairs that are due to
chance projections is performed by randomizing the galaxy positions
and repeating the procedure.

\begin{figure}
  \epsscale{1.1}
  \plotone{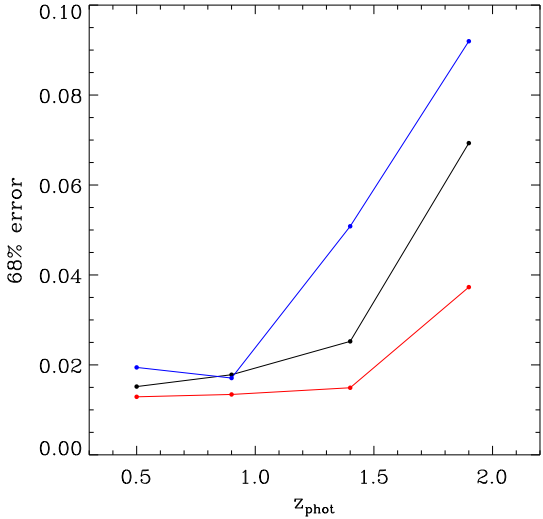}
  \caption{The 68\% photometric redshift errors in
    $\Delta z/(1+z)$ for the mass-selected samples used in this work.
    The errors are estimated by inspecting the difference in
    photometric redshifts of close pairs of galaxies.  The black curve
    is for the full mass-limited sample, and the red and blue curves
    are for quiescent and star-forming subsamples, respectively.}
  \label{fig:zphoterrors}
\end{figure}

The estimated $1\sigma$ photometric redshift errors for the
mass-selected samples used in this work are shown in Figure
\ref{fig:zphoterrors}.  This figure also shows that quiescent galaxies
typically have better photometric redshifts than star-forming
galaxies, which has implications for our density measurements (\S
\ref{sec:density_estimators}).

\subsection{Object selection: stellar masses and the classification of quiescent galaxies}
\label{sec:selection}

Stellar masses were determined by fitting \citet{bruzual03} stellar
population synthesis models to the observed optical/NIR/IR photometry
using FAST \citep{kriek09}\footnote{We note that using the
  \citet{maraston05} models reduces the typical stellar masses by
  $\sim0.15$ dex, with relatively small redshift and mass dependence.}.
  The models were generated using a Chabrier IMF, solar metallicity,
  and a range of exponentially-declining star formation histories.

In this paper we limit the sample at $K_{\rm{AB}}<24.0$, where an
inspection of the galaxy number counts shows that our catalog is
essentially 100\% complete.  This limit corresponds to a $K$-band
signal-to-noise ratio of $\sim8$ in the $1.8\arcsec$ aperture that is
used to measure galaxy colors.  We estimate the mass completeness
limit that corresponds to this flux limit as a function of redshift
using a method similar to that of \citet{marchesini09}: we select
galaxies with $23.5<K<24.0$, scale their fluxes and masses down to our
adopted flux limit of $K=24.0$, and define the mass completeness limit
as the upper end of locus of points in a plot of mass versus
redshift.\footnote{This procedure will fail if galaxies at $K>24$ have
  significantly higher mass-to-light ratios then galaxies at $K<24$,
  as such galaxies would lie above the locus of points if their fluxes
  (and masses) were scaled up to $K=24$.  However this is not expected
  to be the case, as, generally speaking, fainter and lower mass
  galaxies tend to have lower M/L over a wide range in redshift.  We
  have tested our procedure by calculating the mass limit corresponding
  to $K=23$ by selecting galaxies at $22.5<K<23$ and at $23<K<23.5$,
  scaling the fluxes and masses to $K=23$, and verifying that these
  two samples give virtually identical mass completeness limits.}

In this work we select quiescent galaxies using the observed
bimodality in a rest-frame $U-V$ versus $V-J$ color-color diagram.
This has a significant advantage over the standard color-magnitude
diagram that is used to isolate the red sequence in that it
successfully separates out dusty star-forming galaxies from those that
have suppressed star formation \citep{williams09}.  And since it
is based on the observed bimodality, rather than actual estimates of
star formation rates (from e.g.~stellar population modeling), it is
model-independent.  For more discussion, including examples of $U-V$
versus $V-J$ color-color diagrams, we refer the reader to
\citet{williams09} and \citet{whitaker10}.

For brevity we refer to the galaxies selected in such a diagram as
``quiescent,'' although it is possible that they actually have a
non-negligible amount of star formation.  Currently the most stringent
upper limit for such ``quiescent'' galaxies at $z \sim 2$ comes from
the ultradeep NIR spectroscopy of \citet{kriek09}, who quote a star
formation rate of $<4 M_\odot/\rm{yr}$ (corresponding to a specific
star formation rate of $< 2\times10^{-11}\rm{yr}^{-1}$), although
this is only for a single object.  While a galaxy with this much
activity would certainly be considered to be star-forming at $z \sim
0$, \emph{typical} massive star-forming galaxies at these redshifts
have at least an order magnitude more star formation.

\section{Density Estimation, and the Use of Photometric Redshifts}
\label{sec:density_estimators}

Two of the simplest, and traditionally the most widely-used,
estimators of the local projected density are the distance to the
$n$th nearest neighbor, with $n$ typically varying from 3--10, and the
counts within a cylinder, where the radius typically varies from
0.5--8 Mpc.  In this work we use the distance to the $n$th nearest
neighbor, as the adaptive nature of this estimator gives somewhat more
dynamic range in extreme environments \citep[e.g.][]{kovac10}.
Although we present results for mass-selected samples, we do not apply
a mass limit to the ``neighbors'' that are used when estimating the
local densities.  This gives a much higher number of objects on the
sky to estimate the densities with and also allows us to measure
accurate densities over smaller angular scales, thereby limiting the
number of objects that we discard due to edge effects.

In practice we use the distance to the 8th nearest neighbor, which
corresponds to roughly $\sim0.5$ comoving Mpc over the considered redshift
range, but we note that varying $n$ between 5 and 10 yields very
similar results.  We also obtain similar results if counting only
mass-limited objects as ``neighbors,'' although in this case the
length scales that are probed are larger and we have to reduce $n$ to
3--5 in order to limit contamination from objects that are not
physically associated with each other.  Finally, qualitatively similar
results are also obtained by measuring densities within a cylinder
with radius 1 comoving Mpc.  The good (qualitative) agreement between
all of these estimators supports the robustness of our conclusions.

Because the photometric redshifts uncertainties are large when
compared to the length scales of structures in the universe and when
compared to the true virial motions of objects in groups and clusters,
density estimates around individual objects are highly uncertain.
Therefore if even a weak trend is apparent in the data, in actuality
it must be strong indeed.  In the following sections we show
correlations between (projected) local density and galaxy properties,
which suggests that density estimates using broadband photometric
redshifts are still useful and that interesting lessons can still be
learned.  If these estimates were essentially random numbers then we
would not see correlations.

However, photometric redshift errors can actually \emph{introduce}
artificial trends of galaxy properties with density.  This is due to
the fact that quiescent galaxies tend to have more accurate
photometric redshifts than star-forming galaxies \citep[\S
  \ref{sec:zphots}; see also][]{quadri10}.  For example, consider the
case of a galaxy overdensity that has the same quiescent fraction as
the field, so that no SF-density relation exists.  Photometric
redshift errors may scatter some fraction of the star-forming galaxies
in this overdensity to much higher or lower redshifts, in which case
an observer may not consider them to be members of the overdensity at
all.  Thus the observer would infer that the overdensity is populated
primarily by quiescent galaxies, and will mistakenly conclude that an
SF-density relation does exist.

The key issue in this example is what difference in photometric
redshifts can be tolerated before galaxies will no longer be
considered neighbors.  If a sufficiently large redshift ``linking
length'' is used, then the star-forming galaxies in the above example
would still be considered as possible members of the overdensity, and
the hypothetical observer will not be misled.  On the other hand, it
is desirable not to use too large of a linking length in order to
limit contamination from sources that are not physically associated
with each other.

An appropriate value for the linking length would be a few times
larger than the typical photometric redshift uncertainties for the
star-forming galaxies.  This makes it important to have an
understanding of how the redshift errors depend on galaxy type and on
redshift.  As described in \S \ref{sec:zphots} we use the method of
\citet{quadri10} to estimate the uncertainties.  In practice we use a
linking length $3\sigma$.  Varying this between $2\sigma$ and
$4\sigma$ yields very similar results; setting a linking length below
$\sim2\sigma$ makes for a stronger SF-density relation, but this
additional signal may be partially spurious as per the discussion
above.  Additionally, we have tested our procedure by adding noise to
the redshifts of the quiescent galaxies in order to match the
uncertainties for the star-forming galaxies, and our basic conclusions
remain unchanged.

A related issue concerns galaxies that fall outside of our redshift
bins.  When estimating densities for objects near the edge of our
redshift bins, we must be sure to also count neighbors if they fall
outside of the redshift bin of interest but still within a linking
length of the bin.  We have found that neglecting this step can
introduce a strong artificial signal.

Another way in which artificial correlations of quiescent fraction
with density can arise is through the necessity of using galaxy
samples that span a range in redshift.  Since the quiescent fraction
increases with decreasing redshift, and since there is also a slight
trend of increasing density with decreasing redshift within our
redshift windows, this can introduce an artificial trend of increasing
quiescent fraction with increasing density.  In practice this does
have a minor but noticeable effect on our results, so we subtract out
the mean trend of density with redshift within each of our redshift
bins.

\section{The Star Formation-Density Relation}
\label{sec:rfrac}

\subsection{Dependence of Quiescent Fraction on Density and Redshift}

\begin{figure*}
  \epsscale{1.1}
  \plotone{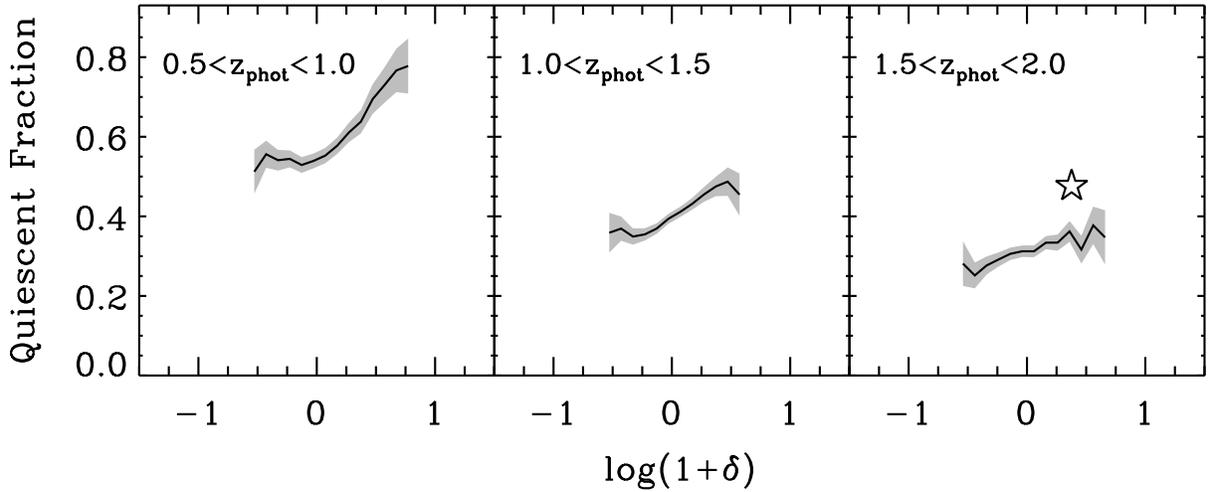}
  \caption{The quiescent fraction versus projected environmental
    density in three different redshift ranges for mass-selected
    samples.  A limit of $\log{(M/M_\odot)} > 10.2$, which corresponds
    to our mass completeness limit at $z = 2$, is applied in all three
    panels.  The quiescent fractions are determined using a running
    mean in a box that is 0.3 dex in $\log{(1+\delta)}$, and the
    shaded region indicates the Poisson uncertainties.  Even given the
    significant uncertainties introduced by photometric redshift
    errors, a clear SF-density relation is found at each redshift.
    The star in the rightmost panel shows the typical density and the
    quiescent fraction of galaxies in the $z=1.6$ cluster described in
    \S \ref{sec:cluster}.  The cluster galaxies have a significantly
    higher quiescent fraction than the rest of the field.}
  \label{fig:red_frac_run_dens}
\end{figure*}

We begin by showing the relationship between quiescent fraction and
projected density in Figure \ref{fig:red_frac_run_dens}.  We apply a
mass limit of $\log{(M/M_\odot)} > 10.2$ in all three redshift bins;
this corresponds to our completeness limit at $z=2$.  Two galaxies are
considered ``neighbors'' if they have a photometric redshift
separation that is $3 \times \sqrt2$ times the estimated redshift
uncertainty, where the factor of $\sqrt2$ accounts for the fact that
both galaxies are subject to redshift errors.  The quiescent fractions
in this figure are calculated using a running mean, where the width of
the box is 0.3 dex in $\log{(1+\delta)}$, and the shaded region
illustrates the $1\sigma$ uncertainties based on Poisson statistics
for the number of quiescent galaxies in a bin.

In agreement with many previous studies, we find a strong star
formation-density relation out to $z \sim 1$, but we also find that
this trend continues to higher redshifts.  In each redshift bin we use
a Kolmogrov-Smirnov (K-S) test to calculate the probability that the
distribution of densities for the quiescent and star-forming
populations are drawn from the same parent distribution.  This
probability is $\ll 1\%$ in the two lower-redshift bins, but increases
to $5\%$ at $1.5<z_{phot}<2$, suggesting that the presence of a
SF-density relation is still significant at the $\sim 2\sigma$ level
at these redshifts.  We also show the typical local density and
quiescent fraction for candidate members of a $z=1.6$ cluster in the
rightmost panel in this figure; this cluster will be discussed in \S
\ref{sec:cluster}.  The cluster candidates have high densities as
estimated using the 8th nearest-neighbor statistic, and also have a
significantly elevated quiescent fraction relative to the field.  Thus
this cluster reinforces our conclusion that the SF-density relation is
in place at $z > 1.5$.

Recently two studies have presented a similar analysis as shown here
using earlier data releases of the UDS.  \citet{tran10} find that the
SF-density relation has reversed by $z \sim 1.6$, whereas
\citet{chuter11} find a normal relation over $1.25<z<1.75$.  Our
analysis is in agreement with the latter study, but the reasons for
the disagreement with \citet{tran10} are unclear.

\subsection{The role of stellar mass in the SF-density relation}
\label{sec:mass}

A physical interpretation of the relationship between quiescent
fraction and density that was shown in the previous subsection is
complicated by the effects of stellar mass.  Since the fraction of
quiescent objects increases with mass, and given the expectation that
more massive galaxies will tend to be found in denser regions, it is
likely that the observed SF-density relation is at least partially due
to a combination of the underlying SF-mass and mass-density relations.
While in the local universe it is well-known that environmental trends
persist even at fixed stellar mass
\citep[e.g.][]{hogg03,baldry06,vandenbosch08,peng10}, the situation is
less clear at higher redshifts: \citet{patel09b} and \citet{cooper10}
find that the same is true at $z \sim 0.9$, but other studies
\citep{scodeggio09, iovino10, grutzbauch11b} have suggested that any
environmental trends at these higher redshifts are solely due to
trends with stellar mass.  If this is true, it may be that these
trends should not be considered to be ``environmental'' at all.

\begin{figure*}
  \epsscale{1.1}
  \plotone{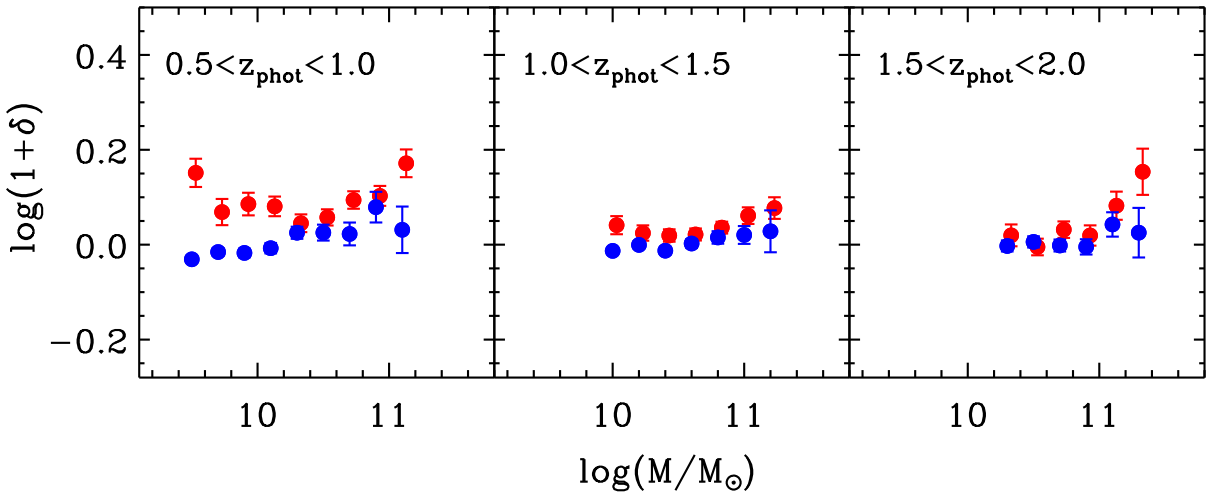}
  \caption{The average density versus stellar mass for star-forming
    galaxies (blue) and quiescent galaxies (red) in 0.2 dex mass bins.
    The red points have been offset slightly to higher masses for
    clarity.  The quiescent galaxies tend to have higher densities
    than the star-forming galaxies even at fixed mass, implying that
    the SF-density relation is not simply the result of an underlying
    mass-density relation combined with a mass-SF relation.  As
    explained in the text, the upturn in the densities for low-mass
    quiescent galaxies suggests that the environment plays a more
    important role in quenching star formation than for higher-mass
    galaxies.  There is also a relationship between stellar mass and
    density at all redshifts, implying that the mass function varies
    with environment.}
  \label{fig:mass_v_density}
\end{figure*}

The relationship between stellar mass and the SF-density relation is
demonstrated in Figure \ref{fig:mass_v_density}, where we show the
mean densities of star-forming and quiescent galaxies in narrow 0.2
dex bins of stellar mass.  Over such narrow bins, the difference in
mean stellar mass between the star-forming and quiescent galaxies is
never more than $0.025$ dex (and is usually much less), which is not
enough to account for a significant change in density.  The error bars
are the standard deviation of the mean.

There are several noteworthy features of this figure.  One is that, as
mentioned above, out to at least $z \sim 1.25$, quiescent galaxies
tend to be found in denser environments than star-forming galaxies
even at fixed mass.  For the star-forming galaxies, there is a clear
and apparently monotonic increase in mean density with stellar mass.
However the relation is different for quiescent galaxies; low-mass
quiescent galaxies are typically found in \emph{denser} regions than
quiescent galaxies of intermediate mass.  This is a strong effect out
to $z \sim 1$, and there is also a hint of it over $1 < z < 1.5$.  A
similar trend of increasing densities at the lowest masses has been
found in the local universe using near-neighbor statistics by
e.g. \citet{hogg03}, and there is also evidence for it in the
correlation functions of \citet{norberg02} and \citet{zehavi05}.  This
result can be understood if the low-mass quiescent galaxies are
primarily satellites in groups or clusters.  Such galaxies would be
forming stars if they were in the field, but the star formation was
shut off sometime after they were accreted into larger systems.
Indeed, \citet{vandenbosch08} use SDSS group catalogs to show that the
majority of quiescent galaxies with $\log{(M/M_\odot)} < 10$ are
satellites, and point to strangulation as the primary physical
mechanism responsible for quenching the star formation.  In the
Appendix we use results from \S \ref{sec:qeff} to demonstrate more
directly how the trends in the left panel of Figure
\ref{fig:mass_v_density} can be understood.

Another feature worth noting in Figure \ref{fig:mass_v_density} is
the simple fact that there is a relationship between stellar mass and
environmental density.  This immediately implies that the mass
function has some environmental dependence, which was already
suggested by the clustering results of \citet{wake11}.

Another way to test whether the observed SF-density relation is simply
due to a stellar mass effect is by creating control samples of star
forming galaxies to compare against the quiescent galaxies.  For each
quiescent galaxy, we find a matching object from among the star
forming sample which has a similar mass and photometric redshift.  We
draw from the star forming galaxies without replacement.  Since
quiescent galaxies dominate the mass function at high masses, there
are not enough star forming galaxies to make even a single complete
control sample.  If we cannot match a quiescent galaxy with a star
forming galaxy, then it is removed.  Conversely, star forming galaxies
strongly dominate at lower masses, so low mass quiescent galaxies have
many possible matches.  In order to recover some of the information
that we lose by not using all of the quiescent galaxies at the massive
end, and not all of the star forming galaxies at the low-mass end, we
repeat this procedure several times in order to create matched sets of
quiescent and star forming galaxies.  Although a more sophisticated
comparison of the star forming and quiescent galaxies is certainly
possible, this method is straightforward and adequate for our purposes.

\begin{figure}
  \epsscale{1.1}
  \plotone{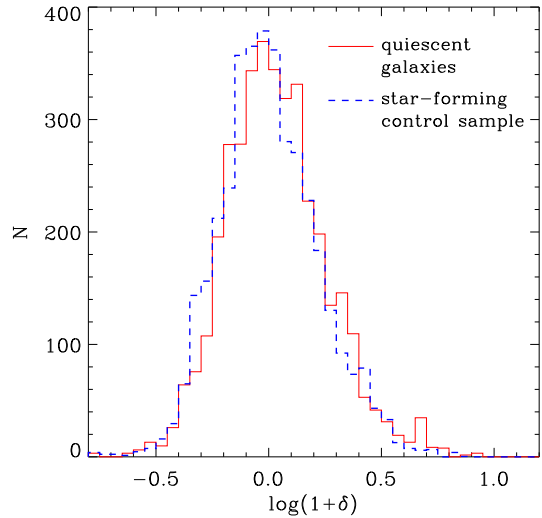}
  \caption{The distribution of densities for quiescent galaxies over
    $1.25<z<1.75$, and for a control sample of star-forming galaxies
    selected to have similar masses and redshifts.  A K-S test
    indicates that the probability that these two distributions are
    drawn from the same parent distibution is $2\%$.}
  \label{fig:ks_histogram}
\end{figure}

Over the redshift ranges considered in Figures
\ref{fig:red_frac_run_dens} and \ref{fig:mass_v_density}, none of the
control samples of star-forming galaxies have mean density that is
higher than the quiescent samples.  A K-S test shows a $\ll 1\%$
chance that the distribution of densities for the quiescent
galaxies and the matched star-forming galaxies are drawn from the same
distribution in our lower redshift bins.  This quantity increases to
2\% at $1.25<z_{phot}<1.75$, suggesting that even at $z \sim 1.5$,
environmental effects can be discerned at fixed stellar mass (see
Figure \ref{fig:ks_histogram}).  However at $1.5<z_{phot}<2.$ this
probability increases further to 24\%, which means that if there is an
environmental effect at fixed mass at over these redshifts then we are
not able to detect it with significance.

\subsection{The relationship between quenching efficiency and stellar mass}
\label{sec:qeff}

Given that the galaxies of all masses are subject to an SF-density
relation out to at least $z \sim 1.5$, it is interesting to consider
whether the environment affects different galaxies in different ways.
In particular, it may be expected that the low-mass galaxies are
affected more strongly than high-mass galaxies.  This does not appear
to be true at lower redshifts: \citet{vandenbosch08} and
\citet{peng10} show that the strength of environmental effects are
independent of stellar mass at $z \sim 0$ \citep[see
  also][]{baldry06}.  \citet{peng10} also show this at $z \sim 0.5$
(see their Figure 7), and suggest that it remains true at even higher
redshifts.

\begin{figure}
  \epsscale{1.1}
  \plotone{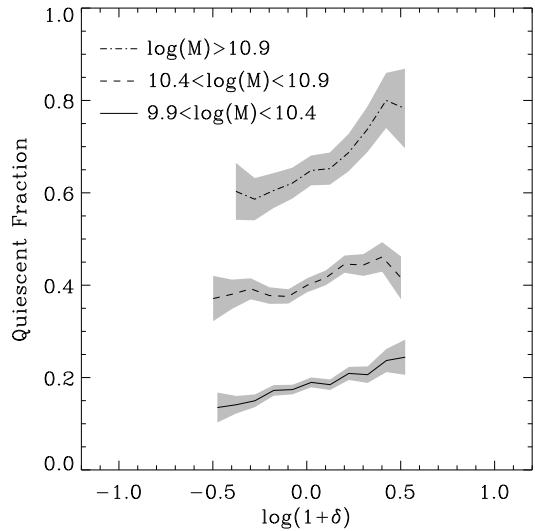}
  \caption{The quiescent fraction versus environmental density at
    $1<z<1.5$ for three different ranges in stellar mass.  An
    SF-density relation is present in each of the mass bins.}
  \label{fig:red_frac_run_dens_masses}
\end{figure}

A first illustration of the effects of environment as a function of
stellar mass is shown in Figure \ref{fig:red_frac_run_dens_masses}.
If the strength of the environmental effects decreases strongly with
increasing stellar mass, it is not apparent in this figure.

\begin{figure*}
  \epsscale{1.1}
  \plottwo{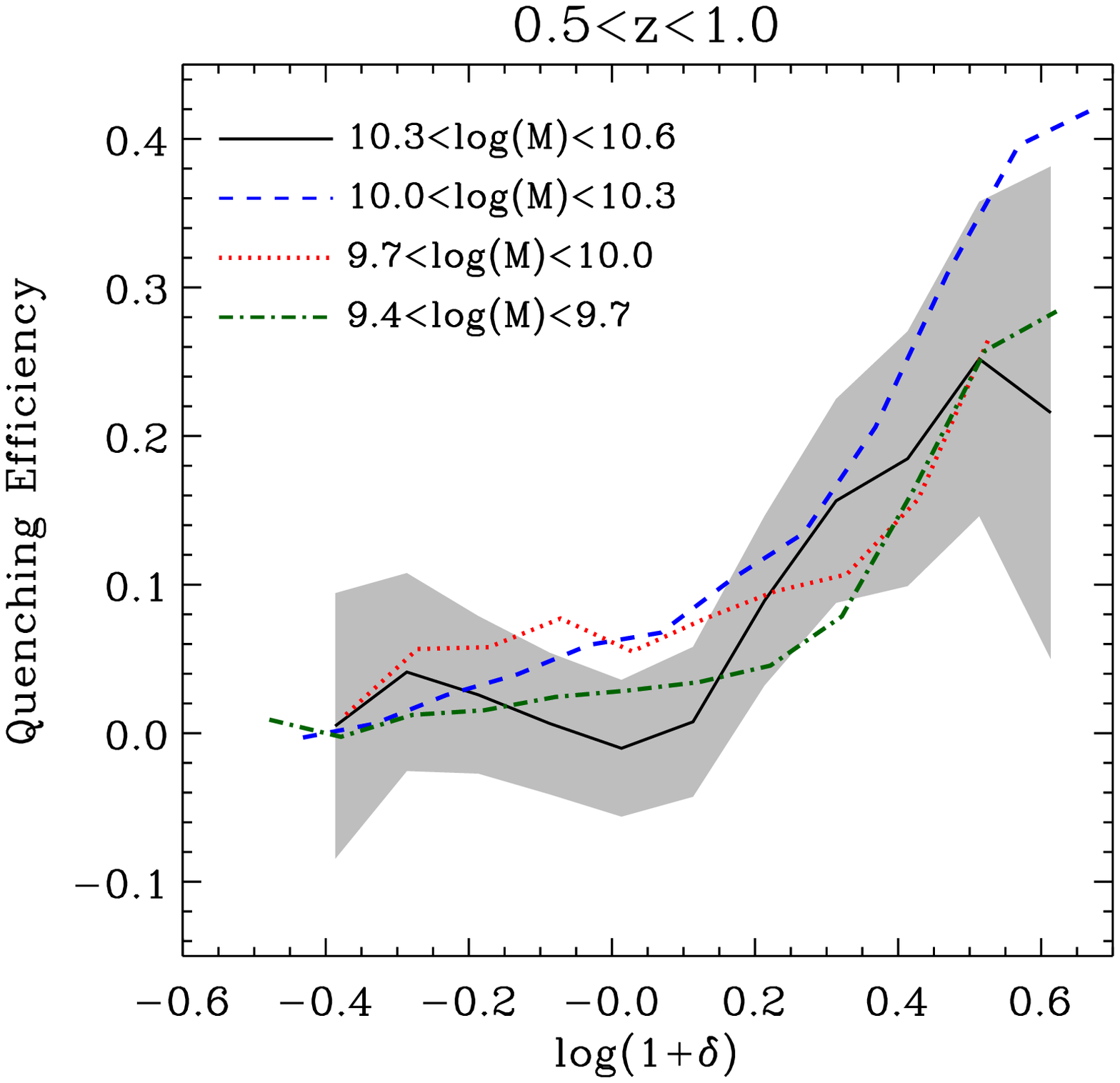}{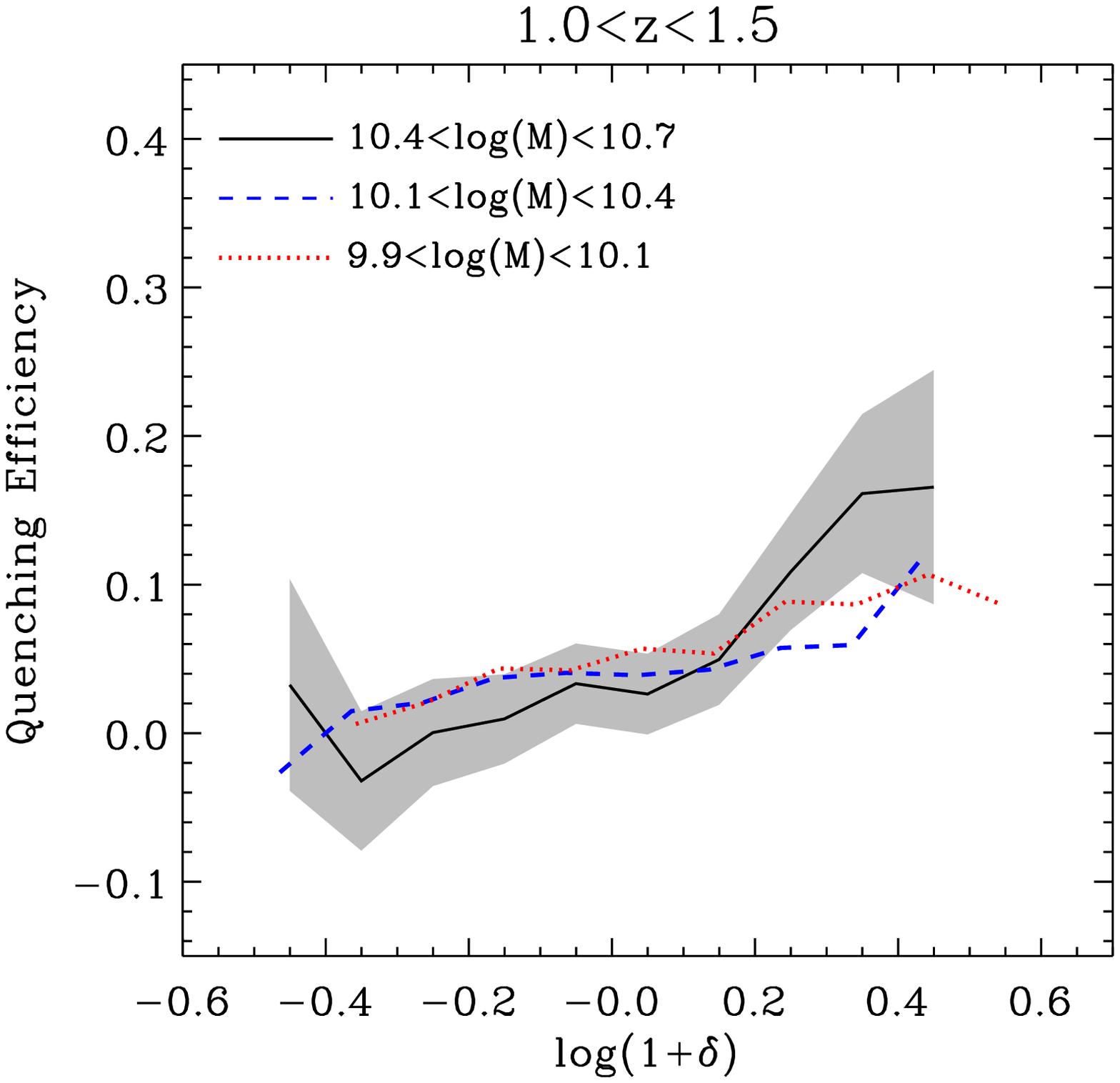}
  \caption{The quenching efficiency versus local density.  The
    quenching efficiency is defined as the fraction of galaxies that
    would be forming stars if they were found in a low-density
    environment, but have had their star formation quenched.  The
    shaded region shows the uncertainties for only the high-mass bins.
    The quenching efficiency appears to be largely independent of
    stellar mass.}
  \label{fig:qeff}
\end{figure*}

However, it is more useful to inspect what we term the
\emph{environmental quenching efficiency}, which is the fraction of
galaxies that would be star-forming if they were in low-density
environments, but have had their star formation quenched due to some
process related to the environment \citep{vandenbosch08,peng10}.  This
is calculated as

\begin{equation}
\varepsilon_q = \frac{f_q(1+\delta) - f_q(1+\delta_0)} {f_{sf}(1+\delta_0)},
\label{eq:qeff}
\end{equation}

\noindent where $f_q$ is the quiescent fraction, $f_{sf} = 1-f_q$ is
the star-forming fraction, and $(1+\delta_0)$ is some low-density
``reference'' environment.  We use $\log{(1+\delta_0)} = -0.4$; this
choice is somewhat arbitrary, but for our purposes the exact value is
not very important.  Figure \ref{fig:qeff} shows the quenching
efficiency in different mass and redshift ranges.  For clarity, we
only show the uncertainties for the highest-mass bin in each panel.
Although the uncertainties in this analysis are significant, it is
apparent that the strength of environmental effect is largely
independent of stellar mass out to at least $z \sim 1.25$, in
agreement with the results from \citet{vandenbosch08} and
\citet{peng10} at lower redshifts.

Figure \ref{fig:qeff} also provides further illustration of the
conclusion that the SF-density relation is not simply due to an
underlying mass-density relation, since at the higher densities all
galaxies---regardless of mass---have a positive quenching efficiency.

\section{A Cluster at z=1.6}
\label{sec:cluster}

In the previous section we showed that the SF-density relation can be
traced to at least $z \sim 1.5$ at fixed mass using a straightforward
statistical comparison of the densities of star-forming and quiescent
galaxies.  Here, we provide a ``case study'' of a $z=1.62$ galaxy
cluster in the UDS field.  This cluster was independently identified
by \citet{papovich10} and by \citet{tanaka10}, and has been detected
in deep \emph{XMM} imaging \citep[even after removal of point
  sources;][]{tanaka10}, making it currently the second
highest-redshift cluster with an x-ray detection.  We follow
\citet{tran10} in choosing the galaxy located at (2:18:21.09,
-5:10:33.1) as the center of the cluster, as it is near the centroid
of the galaxy overdensity, near the peak of the x-ray emission, and is
the brightest cluster candidate in the NIR and \emph{IRAC} bands.

\begin{figure}
  \epsscale{1.1}
  \plotone{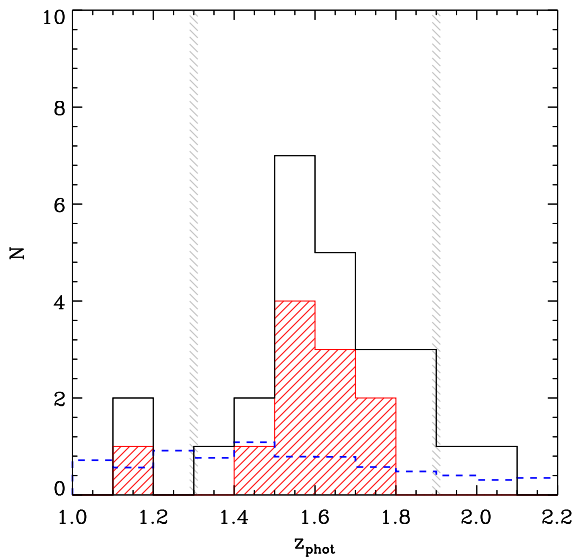}
  \caption{Photometric redshift distribution of objects in the
    vicinity of the cluster at $z=1.62$.  The solid black histogram is
    for all objects that have a stellar mass $M>10^{10} M_\odot$ at
    the best-fitting photometric redshift, and the red hatched
    histogram shows which of those objects are classified as
    quiescent.  The blue dashed histogram shows the expected
    contamination from field galaxies.  The grey vertical hatched
    regions show the photometric redshift range that we use to
    identify candidate cluster members.}
  \label{fig:zhist}
\end{figure}

We select cluster candidates as objects that lie within a projected
distance of 1 comoving Mpc from the cluster center, which is roughly
the radius containing the primary overdensity.  Figure \ref{fig:zhist}
shows the photometric redshift distribution of objects within this
projected distance.  There is a strong peak at the cluster redshift.
As illustrated in the figure, we select as candidates objects at
$1.3<z_{phot}<1.9$\footnote{This redshift range corresponds to a
  photometric redshift error of $\pm 2 \sigma$ at the cluster
  redshift, which means that in principle some cluster members may
  escape our criteria.  However we have tried increasing the search
  range to $3 \sigma$, and a close inspection of SEDs of the
  additional objects suggests that none of them are actual cluster
  members.  Thus using a broader redshift range would only increase
  the contamination from field galaxies.}.

\begin{figure}
  \epsscale{1.1}
  \plotone{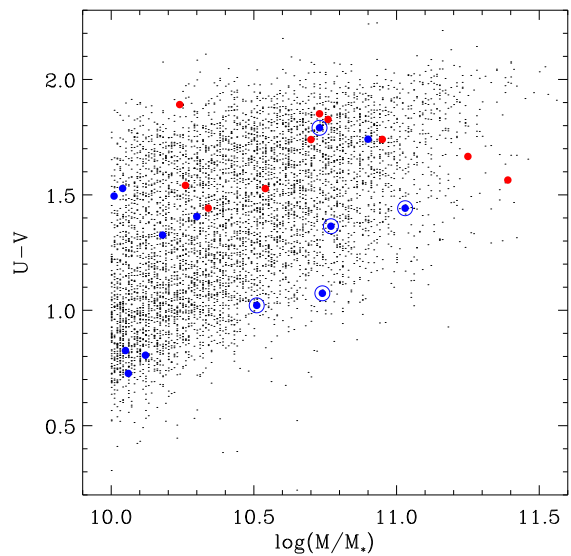}
  \caption{Rest-frame color versus stellar mass for the candidate
    cluster galaxies.  Objects that are classified as quiescent
    according to \S \ref{sec:selection} are shown with red symbols,
    while star-forming galaxies are in blue.  Objects that are
    detected at $24\mu\rm{m}$ are circled.  Quiescent galaxies
    in this cluster are found over the entire mass range, and dominate
    at the highest masses.  At high masses there is good
    correspondence between a $24\mu\rm{m}$ detection and our
    classification of galaxies as star-forming, while quiescent
    galaxies and lower-mass star-forming galaxies are not detected.  The
    small black points represent field galaxies.}
  \label{fig:color_mass}
\end{figure}

As a large majority of the cluster candidates will actually lie in the
cluster, we re-calculate the stellar masses and rest-frame colors of
all candidates after fixing their redshift to $z = 1.62$, and apply
the mass completeness limit of $10^{10} M_\odot$.  Figure
\ref{fig:color_mass} shows a color-mass diagram, with quiescent
galaxies marked in red and star-forming galaxies in blue.  Quiescent
galaxies dominate at $\log(M/M_\odot) > 11$ and star-forming galaxies
dominate at $\log(M/M_\odot) < 10.5$ -- but note that there are
quiescent galaxies even at these relatively low masses.  Figure
\ref{fig:color_mass} also shows those galaxies that are detected at
$>35 \mu\rm{Jy}$ (corresponding to $\sim 3\sigma$) in the \emph{MIPS}
$24\mu\rm{m}$ band.  With a single exception, the more massive
star-forming galaxies are detected while the quiescent galaxies and
the less massive star-forming galaxies are not.  Finally, we point out
that there is no clear bimodality visible in this figure; this is
mostly because several of the star-forming galaxies are highly
reddened, so they contaminate the red sequence \citep[see
  also][]{tran10}.

\begin{figure}
  \epsscale{1.1}
  \plotone{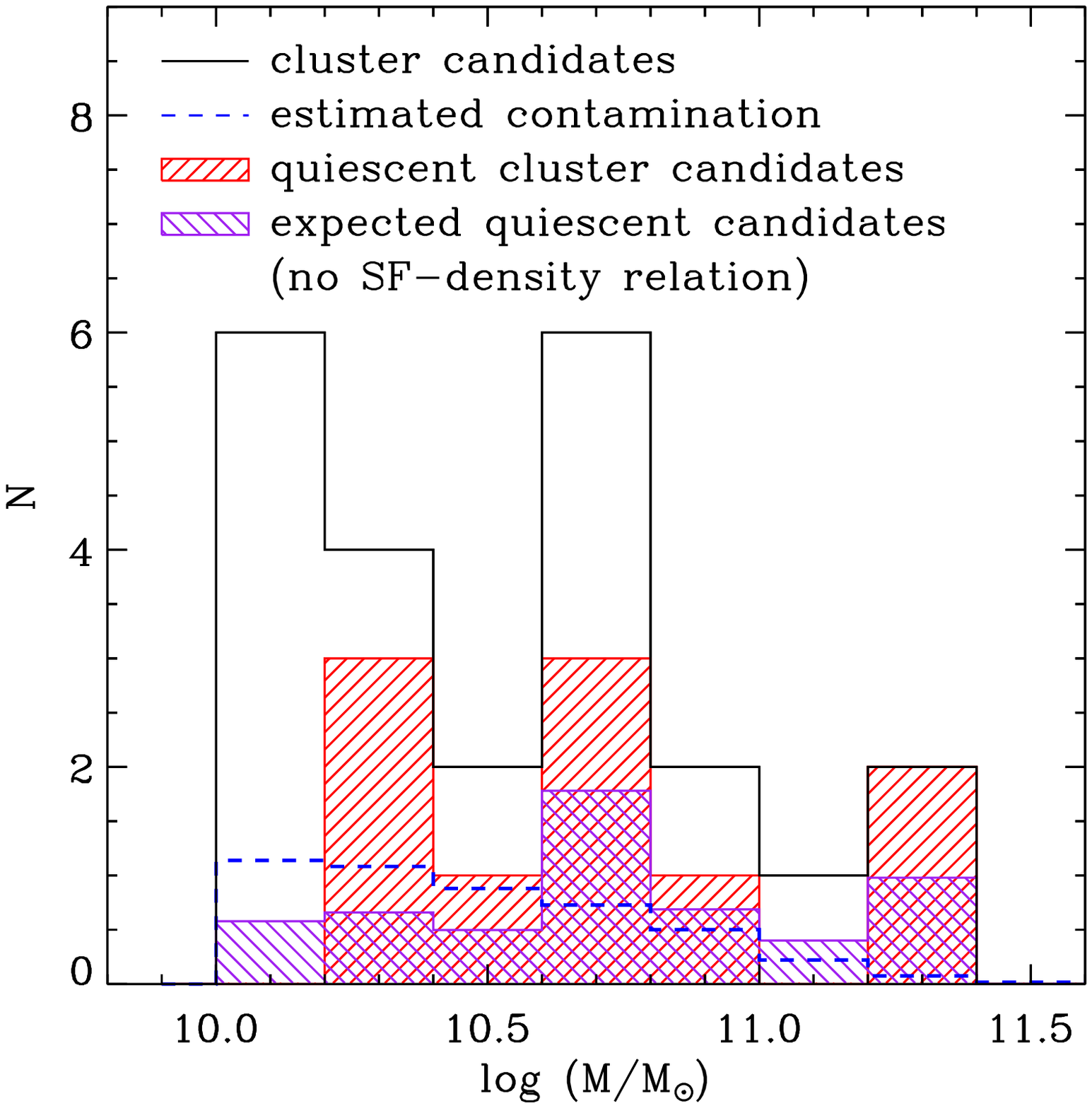}
  \caption{Histograms of stellar masses for candidate members for a
    cluster at $z=1.62$.  The black and red-hatched histograms are for
    the candidates and the subset of quiescent candidates.  The blue
    dashed histogram shows the estimated contamination from field
    galaxies; a comparison with the black histogram shows that
    contamination is minimal.  The purple hatched histogram shows the
    black histogram multiplied by the quiescent fraction as a function
    of mass that is determined from field galaxies; this illustrates
    the expected number of quiescent cluster candidates in the absence
    of an SF-density relation, and a comparison between the red- and
    purple-hatched histograms shows that an SF-density relation is in
    place.}
  \label{fig:mass_hists}
\end{figure}

The masses for the cluster candidates are shown with the black
histogram in Figure \ref{fig:mass_hists}.  There will be some
contamination by field galaxies in our cluster sample, but this should
not be a significant issue considering that this region is overdense
by a factor of 5.  We illustrate the expected contamination with the
dashed blue histogram in Figure \ref{fig:mass_hists}.  The expected
mass distribution of the contaminants was calculated in the same way
as for the cluster candidates: we select all field objects with $1.3 <
z_{phot} < 1.9$, force their redshifts to $z = 1.6$, and re-calculate
the masses.

In Figure \ref{fig:mass_hists} we also compare the number of quiescent
cluster candidates (red hatched histogram) with the number that would
be expected in the absence of an SF-density relation (purple hatched
histogram).  This latter quantity is calculated by multiplying the
mass distribution of all cluster candidates by the mass-dependent
quiescent fraction that is determined from the field.  A comparison of
the hatched histograms shows that the number of quiescent objects (10)
is larger than the expected number of quiescent objects (5.6).  This
immediately suggests the presence of an SF-density relation even at
fixed mass.

\begin{figure}
  \epsscale{1.1}
  \plotone{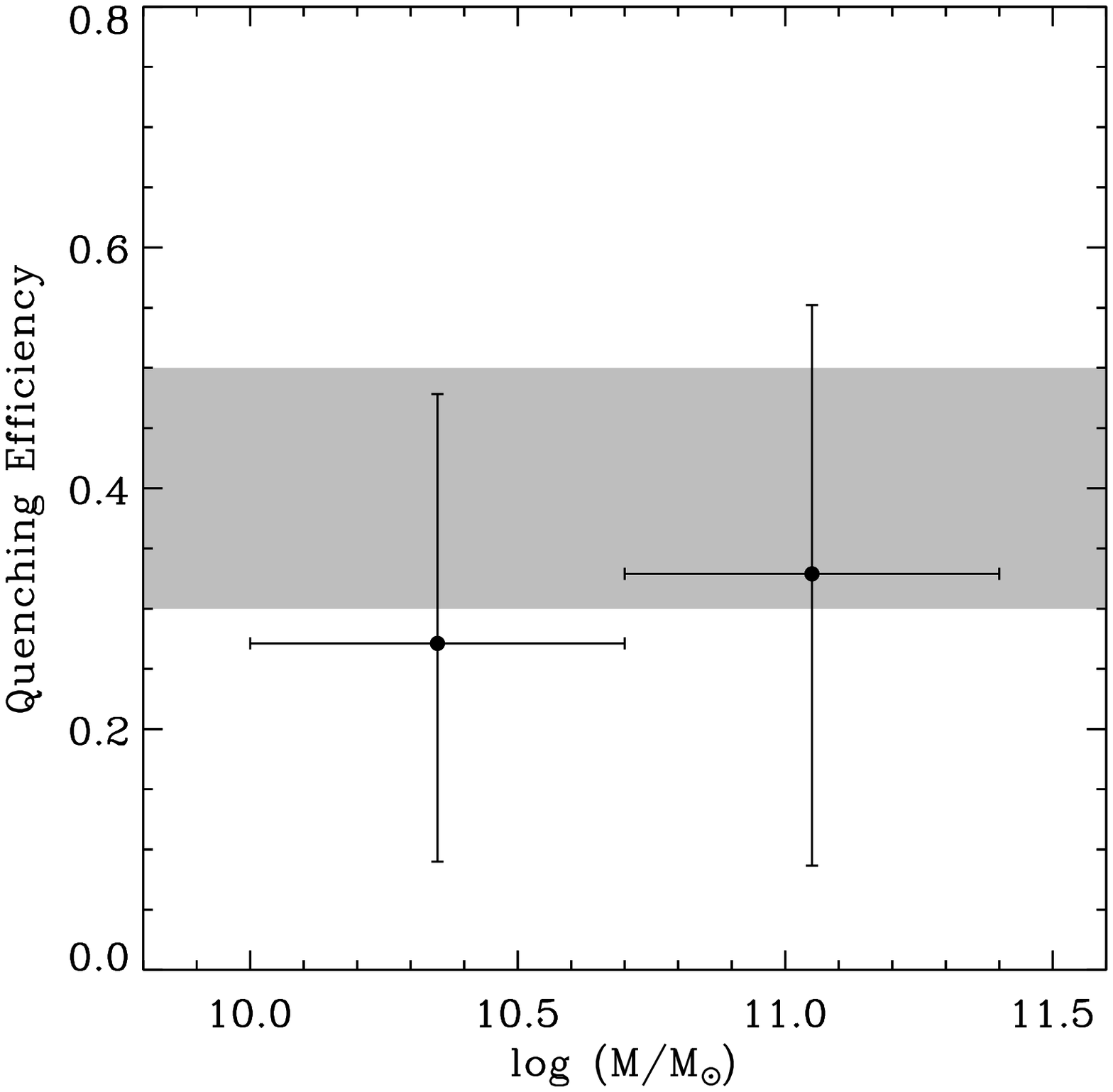}
  \caption{The quenching efficiency in two broad stellar mass bins for
    the candidate cluster members.  The quenching efficiency is
    defined as the fraction of galaxies that would be forming stars if
    found in the field, but have had their star formation quenched.
    Although the uncertainties are large, there is a positive
    quenching efficiency in both mass bins (i.e. there is a SF-density
    relation in this cluster), and there is no evidence that the
    quenching efficiency depends on stellar mass.  The grey shaded
    region shows the range of quenching efficiencies for satellite
    galaxies in SDSS groups from \citet{vandenbosch08}.  Within the
    uncertainties, the quenching efficiency at $z \sim 1.6$ is
    consistent with that at $z \sim 0$.}
  \label{fig:qeff_clust}
\end{figure}

This conclusion is made more explicit in Figure \ref{fig:qeff_clust},
which shows the quenching efficiency for this cluster in two broad
(0.7 dex) mass bins.  The quenching efficiency is calculated in
analogy with equation \ref{eq:qeff}, where the low-density
``reference'' environment is taken to be the field and the
higher-density environment is the cluster.\footnote{Since many of
  these ``field'' galaxies will also be in groups, and will therefore
  have experienced some quenching due to environmental processes, our
  quenching efficiencies will be biased slightly low.} In this figure
we have also performed (small) corrections for the estimated
contamination.  The uncertainties in the quiescent fraction are
calculated using the Wilson interval for binomial statistics, which is
appropriate for the small numbers considered here.

Although the uncertainties are large due to the limited sample size,
there does appear to be a boosted quenching efficiency in each of our
(independent) mass bins.  Also, as was already shown in Figure
\ref{fig:qeff} at somewhat lower redshifts, there is no suggestion
that the quenching efficiency depends on stellar mass.  The grey
shaded region in this figure illustrates the range of quenching
efficiencies determined by \citet{vandenbosch08} for satellite
galaxies in groups in the SDSS.  Although the large uncertainties
obviously prevent us from drawing strong conclusions, it is striking
that the quenching efficiency at $z = 1.6$ is consistent with that at
$z \sim 0$.

We note that \citet{tran10} have also investigated the star formation
in this cluster.  Those authors note that a significant number of the
cluster candidates are detected at $24\mu\rm{m}$, and are therefore
likely to be strongly star-forming.  They also emphasize that this
situation is different than for clusters at $z \lesssim 1$, where a
larger fraction of galaxies are dead.  Here we emphasize a
complimentary point: while it is true that a significant fraction of
the cluster members are forming stars at a high rate, nonetheless the
fraction of quiescent objects is \emph{still} higher than in the field
at similar redshifts.\footnote{There are a number of differences
  between our analysis and that of \citet{tran10}.  The most important
  appears to be that they identify cluster candidates within 1
  physical Mpc, whereas we use 1 comoving Mpc; their larger radius
  will include significantly more contamination from field galaxies,
  and will include more galaxies on the lower-density outskirts of the
  cluster which may be expected to have increased SF activity.  A
  second difference is that we use mass-selected samples.  The fact
  that \citet{tran10} focus on star formation rates derived from
  $24\mu\rm{m}$ imaging, whereas we classify galaxies according to the
  color bimodality, is less material: we find that $\sim 20 \%$ of the
  cluster candidates are detected at $>35 \mu\rm{Jy}$ at
  $24\mu\rm{m}$, whereas this increases to $\sim 40 \%$ over the rest
  of the field.  Despite the differences in methods, \citet{tran10} do
  find a quiescent fraction that is simlar to ours (see the bottom
  panel of their Figure 3).} Thus, a SF-density relation appears to be
in place at $z=1.6$ with no evidence of a reversal.

\section{Summary and Discussion}
\label{sec:summary}

In this paper we have studied the evolution of the star
formation-density relation, in which quiescent galaxies are found
preferentially in dense environments.  We have used public data from
the UKIDSS-UDS and have constructed mass-limited samples, rather than
flux- or luminosity-limited samples, in order to obtain a better
understanding of environmental effects and to avoid the selection
effects that have been prevalent in several previous studies.
Quiescent galaxies are identified according to the observed bimodality
in a rest-frame color-color diagram; we prefer this method over the
use of a standard color-magnitude diagram as it cleanly separates
galaxies that are red due to dust from galaxies that are red due to a
lack of significant star formation.  Environmental densities are
measured by counting near neighbors.  Because large and representative
samples of spectroscopic redshifts are not currently available at
$z>1$, in this work we rely on photometric redshifts.  This introduces
large uncertainties in the density measurements and, as discussed in
\S \ref{sec:data}, can also potentially introduce spurious
environmental trends if the effects of redshift errors are not
correctly taken into account.

Even with the uncertainties inherent in our analysis, we find that the
SF-density relation can be traced to at least $z \sim 1.8$, which is
higher than previous studies have found using either photometric or
spectroscopic redshifts (but see \citealt{chuter11}, who have recently
used earlier data in the UDS to arrive at a similar conclusion).  We
show that, out to at least $z \sim 1.5$, the SF-density relation is
not simply the result of a mass-density relation combined with a
mass-SF relation: even at fixed mass galaxies in denser environments
have a higher quiescent fraction.

Nevertheless we do find a relationship between stellar mass and
environment.  As shown in Figure \ref{fig:mass_v_density}, this
relationship is straightforward for star-forming galaxies in that more
massive galaxies tend to be found in denser environments.  The
situation is more complicated for quiescent galaxies: both low-mass
($\log(M/M_\odot) \lesssim 10.3$) and high-mass quiescent galaxies
tend to be found in the densest environments, whereas at intermediate
masses galaxies are also found at lower densities.  This is a known
feature of the low-redshift universe (\citealt{hogg03};
\citealt{vandenbosch08}; see also \citealt{ross10}), and we have found
that it remains in place out to at least $z \sim 1$.  As we
demonstrate in the Appendix, this relationship can be understood if
low-mass quiescent galaxies occur primarily at high densities because
environmental processes are required to shut off the star formation.
At intermediate masses there are other processes (e.g.~AGN feedback)
that can also play a role; such galaxies may thus also be found in
comparatively low-density environments.  At even higher masses
galaxies tend to be found in groups and clusters regardless of whether
they are forming stars.  This reasoning implies that the environment
plays a greater role in the build-up of the red sequence at lower
masses \citep[Fig. \ref{fig:model_qeff}; see also][]{vandenbosch08}.
It also implies that the shape of the mass function has significant
environmental dependence at $z \sim 1$.  For the star-forming
galaxies, we expect that dense environments should have a greater
number of massive galaxies when compared to the field.  For the
passive galaxies, we expect that dense environments have a greater
number of high-mass \emph{and} low-mass galaxies.

We find that environmental quenching operates on all galaxies, with no
evidence that the quenching efficiency depends on stellar mass.
Similar results have been obtained at lower redshifts by
\citet{vandenbosch08} and \citet{,peng10}.

Most of our conclusions receive additional support from our analysis
of the central regions of a cluster at $z = 1.62$.  The
photometrically-selected candidate cluster members tend to have higher
stellar masses than field galaxies, and have a higher quiescent
fraction even at fixed mass.  Although the uncertainties are very
large, there does not appear any variation in the quenching efficiency
with mass within the cluster.  Interestingly, the quenching efficiency
is also consistent with that found by \citet{vandenbosch08} for
low-redshift galaxy groups and clusters in SDSS.  This is also in
agreement with the suggestion by \citet{peng10}, who find that the
environmental quenching efficiency does not evolve significantly at $z
< 1$.  But, as pointed out by those authors, environmental processes
may still play a greater role at lower redshifts as more galaxies
will have been accreted into dense environments.  In this paper we
have only considered galaxies in the central regions of a single
cluster; it would clearly be beneficial to extend this analysis to a
greater number of high-redshift clusters, and to look for variations
in galaxy properties with clustercentric distance
\citep[e.g.][]{patel09a}.

The results presented here have implications for our understanding of
the physical processes that work to quench star formation in dense
environments.  It is thought that strangulation --- the stripping of
halo gas from galaxies as they are accreted into larger systems --- is
largely responsible, but that this is a gradual process that acts over
the course of a couple Gyr
\citep[e.g][]{mccarthy08,mcgee09,weinmann10}.  If this remains the
case at high redshift then it is expected that environmental effects
should cease to be observable since satellite systems can only have
been accreted recently.  Indeed, \citet{mcgee09} predict that
environmental trends should be weak or non-existent at $z \sim 1.5$.
Even with the large uncertainties introduced by our use of photometric
redshifts, we do find an SF-density relation in place at this
redshift.  It may be that the relevant timescales (for gas stripping,
and for the consumption of gas that hasn't been stripped) are shorter
at these redshifts \citep{weinmann10,tinkerwetzel10}.  Other physical
processes that operate in dense environments may also play a role.
Another slightly more exotic possibility is that, one way or another,
the galaxies ``knew'' beforehand that they would be accreted and had
already shut off their star formation; such an effect would presumably
be related to the ``assembly bias'' of dark matter halos \citep[see
  discussion by][]{quadri08,tinker10,neistein10}.  Unfortunately we
cannot detect nor completely rule out a relation at significantly
higher redshifts to provide tighter constraints.

Our result that the environmental quenching efficiency does not show a
strong dependence on stellar mass --- which echoes recent results at
lower redshifts --- may present an interesting challenge, as
strangulation is expected to be somewhat more effective for low-mass
galaxies \citep[e.g.][]{mccarthy08}.  Although the uncertainties are
large, our analysis of the $z=1.6$ cluster suggests that the
environmental quenching efficiency is not greatly reduced compared to
at $z \sim 0$.  This may also present interesting constraints, as
galaxies in this cluster can only have been accreted recently so
environmental processes have had a much longer time to operate at $z
\sim 0$ than at $z \sim 1.6$.

In this paper we have investigated the fraction of quiescent objects
as a function of local density.  We note that, at least in principle,
it is possible that other measures of environmental influence --- such
as the average star formation rate per galaxy or the fraction of
galaxies undergoing intense starbursts --- would yield different
results.  Similarly, the color-density relation may evolve differently
since many ``red'' galaxies have obscured star formation, and the
morphology-density relation may also evolve differently
\citep{capak07}.  Future work would do well to investigate each of
these relationships.

The analysis that has been presented in this paper is quite basic, and
can be extended and improved upon in several ways.  It may not be
completely straightforward to significantly improve on the simplest
density estimators similar to the one used in this paper using
standard broadband photometric redshifts, and obtaining very large and
unbiased samples of spectroscopic redshifts beyond $z \sim 1$ remains
difficult.  More sophisticated means of constructing the density field
that include both photometric and spectroscopic redshifts provide one
direction forward \citep{kovac10}.  Another difficulty lies in a
detailed physical interpretation of environmental densities.  In the
$\Lambda$CDM context, it is not necessarily the number of near
neighbors that is the most relevant physical quantity; more relevant
``observables'' include halo mass, whether a galaxy is a central or
satellite, and perhaps group- or cluster- centric distance.
Nearest-neighbor statistics mix all of these quantities.  Thus it may
be more promising to identify groups, and to distinguish between
centrals and satellites in the same way that has been done at lower
redshifts.  However this procedure is also problematic, as identifying
group members is not trivial and, as groups should be rapidly
assembling at these redshifts, the central/satellite distinction may
be less meaningful.

\acknowledgements 

This work is based on data obtained as part of the UKIRT Infrared Deep
Sky Survey.  We thank Gabe Brammer for help with the photometric
redshifts, as well as Simone Weinmann, Shannon Patel, and Olivera
Rakic for their careful reading of a draft version of this manuscript.
Support for this work was provided by NASA through Hubble Fellowship
grant \#51279.01 awarded by the Space Telescope Science Institute,
which is operated by the Association of Universities for Research in
Astronomy, Inc., for NASA, under contract NAS 5-26555.

\appendix
\section{The Different Mass-Density Relations for Star-Forming and Quiescent Galaxies, and Implications for the Build-Up of the Red Sequence}
\label{sec:appendix}

The leftmost panel in Figure \ref{fig:mass_v_density} shows a roughly
monotonic increasing relationship between mean density and stellar
mass for star-forming galaxies.  But for quiescent galaxies the
relationship is less trivial, since both low-mass and high-mass
quiescent galaxies lie in denser regions than those of intermediate
mass.  This may seem counterintuitive, since low-mass galaxies are not
expected to be particularly associated with high-density regions.  In
this appendix we use a simple model to show how these relationships
might be understood, and to support our conclusion that --- even
though the efficiency of environmental quenching appears to be
independent of stellar mass --- the environment plays a larger role in
building up the red sequence at lower masses than at higher masses.
The treatment here is only meant to be illustrative and approximate,
as a complete treatment would require better three-dimensional density
estimates over a wide dynamic range than is possible using our simple
and purely photometric projected density measurements.

\begin{figure*}
  \plottwo{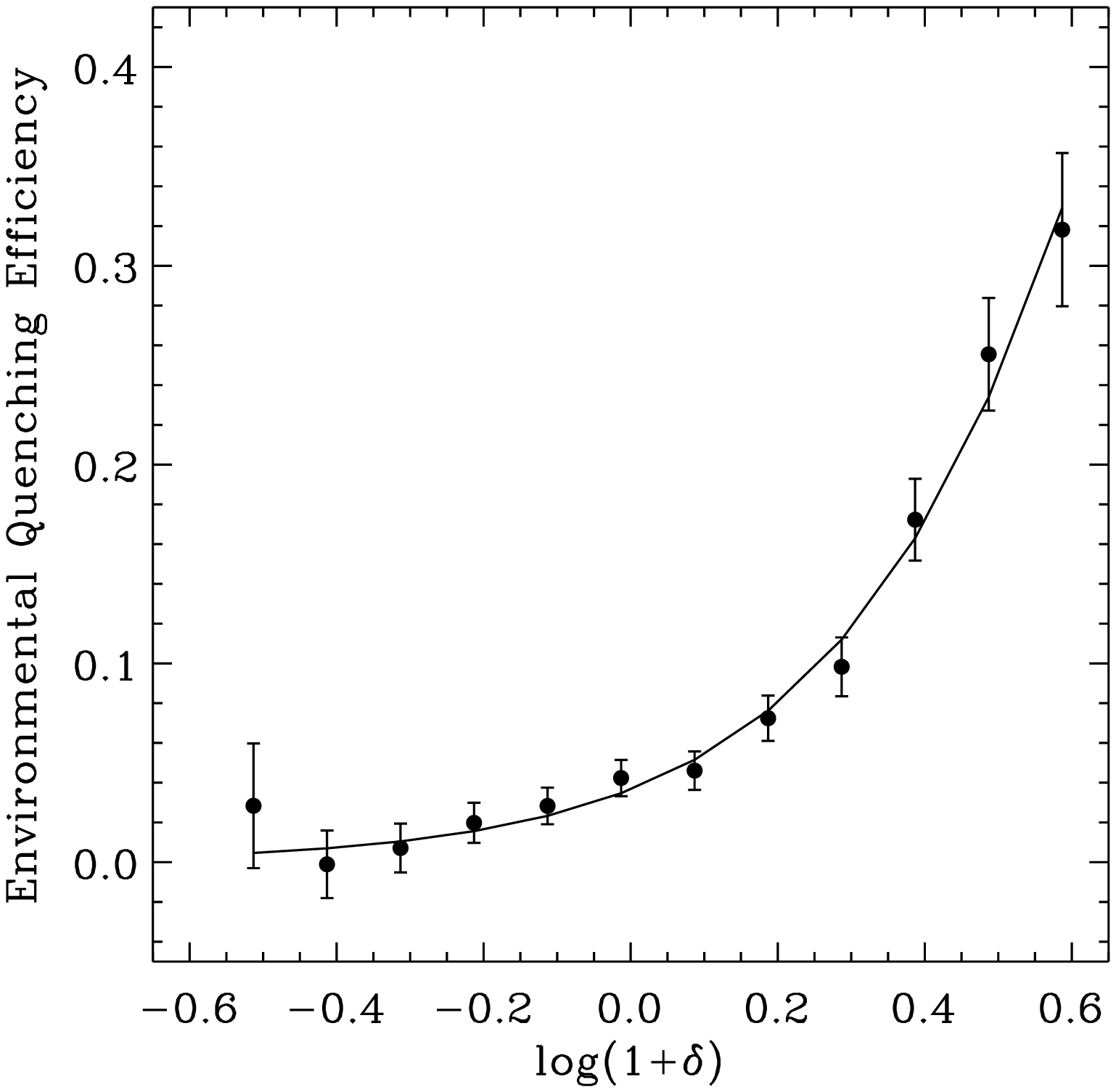}{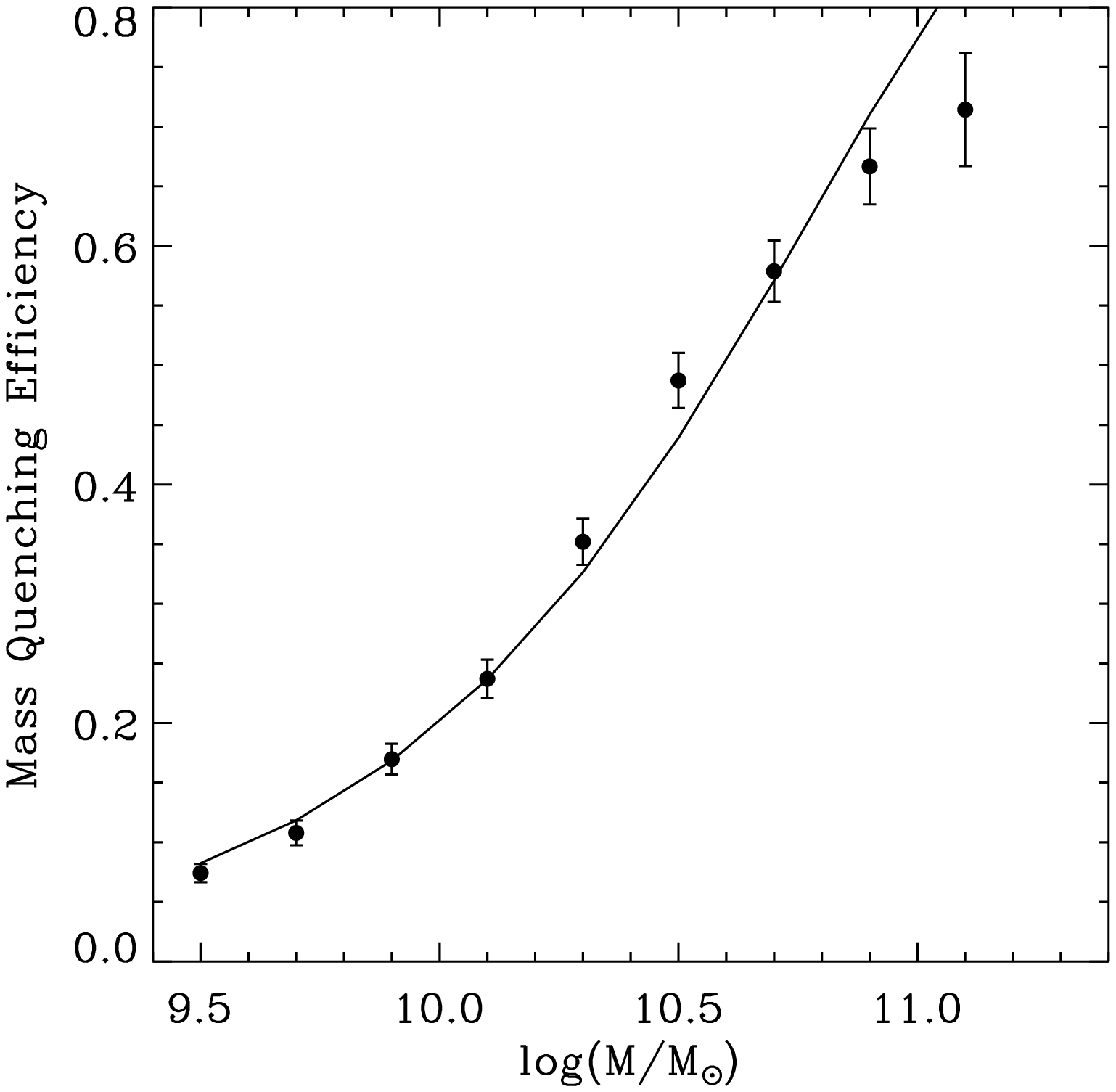}
  \caption{\emph{Left:} The environmental quenching efficiency, which
    is defined as the fraction of galaxies that would be forming stars
    if they were found in low-density environments, but have had their
    star formation shut off by environmental processes.  \emph{Right:}
    The mass quenching efficiency, which is the residual quenching
    that is not explained by environmental quenching.  This quenching
    process is a strong function of stellar mass.  The solid curves in
    the two panels are fits to the data.}
  \label{fig:fit_qeff}
\end{figure*}

Our approach is similar to that taken previously by \citet{peng10},
and involves three basic ingredients.  The first is a mass-density
relation, in which more massive galaxies tend to lie in denser
regions.  The second is the environmental quenching, which we
calculate as described in \S \ref{sec:qeff}.  However there must be
(at least) one other quenching mechanism.  We will refer to this third
ingredient as \emph{mass quenching} as it is found to be a strong
function of stellar mass.  Roughly speaking, this results in the
well-observed phenomenon that more massive galaxies are more likely to
be quenched.  We calculate the mass quenching from the fraction of
quenched objects which are not explained by environmental quenching
over the redshift range $0.5<z_{phot}<1$.  The environmental and mass
quenching efficiencies are shown in the two panels of Figure
\ref{fig:fit_qeff}, and in both cases we fit a function of the form

\begin{equation}
\epsilon(\rho) = 1 - \exp((-\rho/p_1)^{p_2}).
\label{eq:qeff_func}
\end{equation}

\begin{figure}
  \plotone{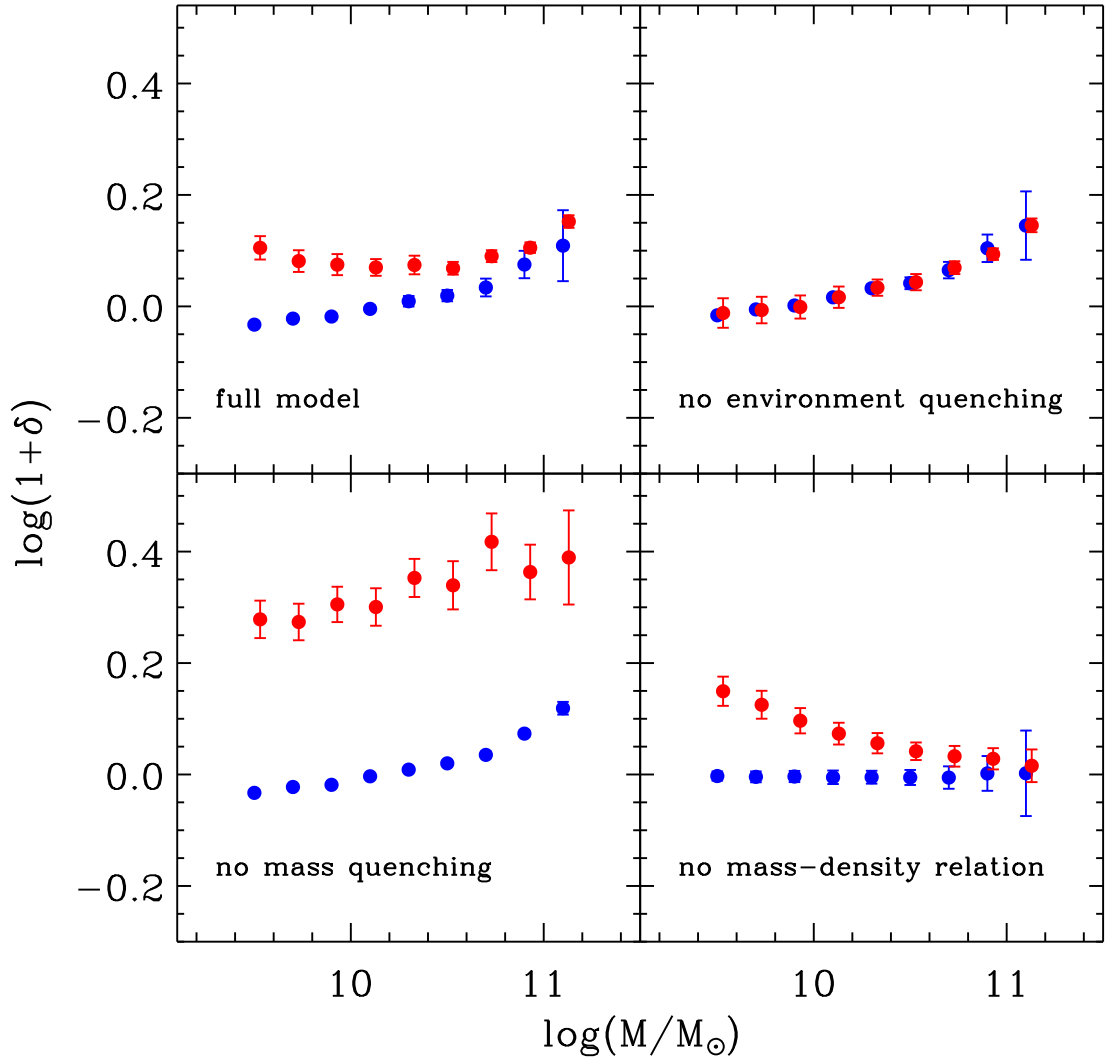}
  \caption{The top left panel shows the relationship between stellar
    mass and mean density for quiescent and star-forming galaxies (red
    and blue symbols, respectively) in our Monte Carlo simulations.
    The errorbars represent the scatter between the simulations, and
    the red symbols have been offset slightly to the right for
    clarity.  This panel can be compared to the measurements shown in
    the leftmost panel of Figure \ref{fig:mass_v_density}.  In the
    remaining three panels of this figure, we remove in turn each of
    the three ingredients in the Monte Carlo simulations.  It is
    apparent that all three ingredients interact to produce the
    characteristic shapes in the top left panel.}
  \label{fig:model_mass_v_density}
\end{figure}

We perform Monte Carlo simulations in which we take the observed
masses and densities for galaxies at $0.5<z_{phot}<1$ and randomly
quench them according to the environmental and mass quenching
efficiencies.  The main result of these simulations is shown in the
top left panel of Figure \ref{fig:model_mass_v_density}.

It is perhaps not surprising that the simulations more or less
reproduce the measurements shown if Figure \ref{fig:mass_v_density},
as the simulations make use of the densities, mass quenching
efficiency, and environmental quenching efficiency that we infer from
the data.  The key point is that the three ingredients described above
are required to produce the characteristic shapes of the mass-density
relations for both the star-forming and quiescent populations.  In the
remaining panels of Figure \ref{fig:model_mass_v_density} we remove
each of the three ingredients in turn in order to provide some insight
into how they interact to produce these characteristic shapes.  In the
top right panel, we remove the environmental quenching; this has the
obvious effect of eliminating the difference in densities between
quenched and star-forming galaxies.  In the bottom left panel we
remove the mass quenching.  In this case it is primarily the galaxies
in the densest regions that are quenched, since mass quenching is not
available to quench galaxies at less extreme densities.  Finally, in
the bottom right panel we remove the mass-density relation by
randomizing the relationship between stellar mass and densities in our
catalogs.  In this case there is no relationship between mass and
density for the star-forming galaxies, and there is no upturn at high
masses for the quiescent galaxies.  But there is a prominent upturn in
the densities for low-mass quiescent objects.  This is because mass
quenching is very inefficient for those objects
(Fig. \ref{fig:fit_qeff}) and so it is primarily the few low mass
galaxies that are in very dense regions that are quenched.  A
consideration of the top right and bottom panels of Figure
\ref{fig:model_mass_v_density} shows how each of the three ingredients
of the model interact to produce the shapes that are shown in the top
left panel: the environmental quenching introduces a large offset
between the star-forming and quiescent tracks, the mass quenching
reduces the offset (doing so more effectively at intermediate and high
masses than at low masses), and the mass-density relation causes the
rise in densities at high masses for both the star-forming and
quiescent galaxies.

\begin{figure}
  \plotone{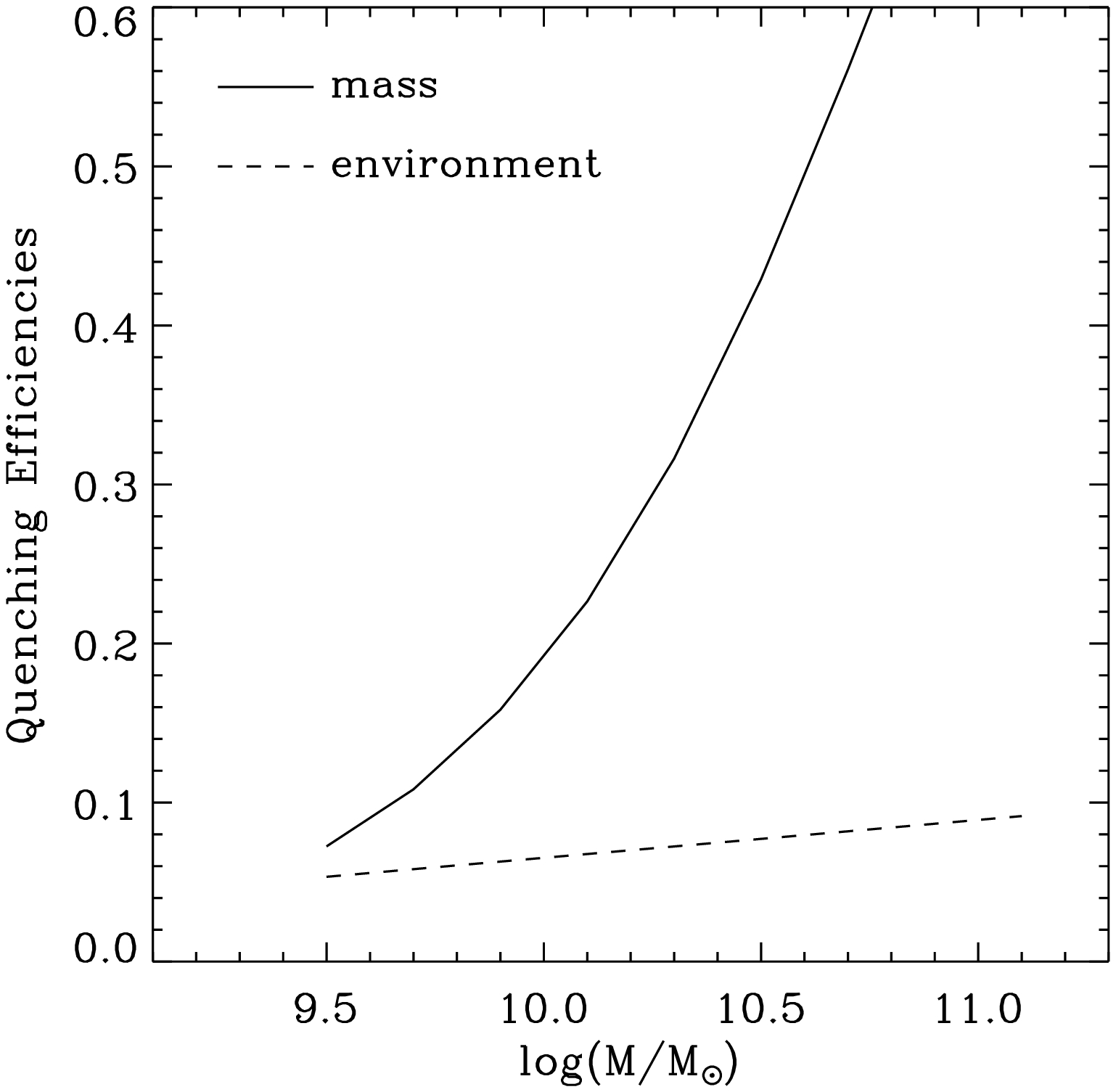}
  \caption{The mass and mean environmental quenching efficiencies as a
    function of stellar mass.  Mass quenching dominates strongly over
    environmental quenching at all but the lowest stellar masses.
    This implies that environmental processes play a relatively small
    role in the buildup of the red sequence at high masses, but
    become increasingly important at lower masses.}
  \label{fig:model_qeff}
\end{figure}

In \S \ref{sec:mass} and \S \ref{sec:summary} it was argued that the
U-shaped curve followed by the quiescent galaxies in the the plot of
mass versus density suggests that the environment plays a more
important role in building up the red sequence at lower masses.  This
is shown explicitly in Figure \ref{fig:model_qeff}, which compares the
mass and environmental quenching efficiencies as a function of stellar
mass.  The environmental quenching efficiency in this figure is
calculated as the mean value for galaxies in bins of stellar mass.  It
is apparent that the mass quenching dominates significantly over the
environmental quenching at all but the lowest stellar masses.  It is
only at these relatively low masses that environmental processes play
a large role in building up the red sequence \citep[see
  also][]{vandenbosch08}.  The fact that the environmental quenching
efficiency is lower than the mass quenching efficiency does not mean
that environment does not exert a strong influence; it merely reflects
the fact that most galaxies are not in dense enough regions for
environmental processes to have a large effect.  In this appendix we
have focused on the redshift range $0.5<z<1$, and it is expected that
the curves shown in Figure \ref{fig:model_qeff} will evolve somewhat
at lower and higher redshifts.

\end{document}